# Anisotropic superconducting properties of MgB$_2$ single crystals probed by in-plane electrical transport measurements.

Yu. Eltsev, S. Lee, K. Nakao, N. Chikumoto, S. Tajima, N. Koshizuka, M. Murakami

Superconductivity Research Laboratory, ISTEC,
10-13, Shinonome 1-chome, Koto-ku, Tokyo, 135-0062, Japan

*We report on study of the electronic anisotropy of the newly discovered MgB$_2$ superconductor using the in-plane resistivity measurements in magnetic field applied perpendicular and parallel to Mg and B planes of MgB$_2$ single crystals. The results show temperature dependent anisotropy of the upper critical field with anisotropy ratio $\gamma = H_{c2//}/H_{c2\perp}$ increasing from 2.2 close to $T_c$ up to about 3 below 30K. Our estimation of the in-plane and out-of-plane coherence length of about $\xi_{ab}(0)=68$Å and $\xi_c(0)=23$Å and electronic mean-free path $l_{ab}=240$Å and $l_c=60$Å respectively indicates MgB$_2$ single crystal as approaching clean limit type-II superconductor.*

PACS numbers: 74.25.Fy, 74.70.Ad, 74.60.Ec

Discovery of superconductivity at about 39K in magnesium diboride[1] has stimulated considerable interest in study of various properties of this compound. Observation of the boron isotope effect[2] as well as band structure calculations[3] indicate phonon-mediated superconductivity in $MgB_2$. Transport and magnetization measurements of the upper critical field[4-8] and critical current density[9,10] of $MgB_2$ revealed properties typical for a type-II superconductor. Since crystal structure of $MgB_2$ consists of alternating Mg and B sheets, electronic anisotropy of this material may be expected. Clear signs of anisotropic nature of $MgB_2$ has been demonstrated from measurements performed on thin films,[11,12] aligned crystallites[13] and fine powder[14] samples. However, results obtained by various groups give very different estimation of $H_{c2}$ anisotropy ratio ranging from =1.7-2 obtained for thin films[11,12] and aligned crystallites[13] up to =6-9 for randomly oriented powder samples.[14]

Recently Lee *et al.* reported on growth of sub-millimeter $MgB_2$ single crystals under high pressure in Mg-B-N system.[15] Availability of such crystals opens a nice opportunity for direct probe of anisotropic behavior of $MgB_2$ superconductor, and from resistivity measurements performed in Ref. 15 the upper critical field anisotropy ratio of about 2.7 was estimated.[16] Here we present results of the detailed study of the in-plane transport properties of $MgB_2$ single crystals in magnetic field up to 6T applied perpendicular to Mg and B planes (H ) and up to 16T in parallel field orientation ($H_{//}$). Obtained results allow to determine the upper critical fields and, thus, to get information about superconducting coherence lengths and superconducting state anisotropy.

Magnesium diboride single crystals have been grown in quasi-ternary $Mg-MgB_2$-BN system at high pressure and temperature of 4-6 GPa and 1400-1700°C respectively. The $MgB_2$ precursor was prepared from magnesium powder (99.9% Rare Metallic Co.) and amorphous boron (97% Hermann C.Starck). Single crystal growth was performed in BN crucibles in the presence of longitudinal temperature gradient of about 200°C/cm in a cubic-anvil press (TRY Engineering). In optimal conditions shiny gold-coloured single crystals of size up to 0.7mm were grown. Several plate-like single crystals of size of about $0.5 \times 0.1 \times 0.05 mm^3$ have been chosen for our study. The in-plane transport measurements have been performed in a usual four probes linear geometry with transport current directed along Mg and B planes. Electrical contacts were made using gold or silver paste without subsequent heat treatment. Contact resistance was around 1  for current contacts and slightly higher (about 3-5 Ω) for potential ones. To measure current-voltage response, we used usual low frequency (17Hz) lock-in ac technique with excitation current in the range 0.2-2.0mA and voltage resolution of about 0.3nV. Magnetic fields up to 9T were generated by superconducting solenoid, while measurements in high fields up to 16T were performed in a pulsed magnet. For measurements inside superconducting solenoid the samples were mounted in a rotatable sample holder with an angular resolution of 0.05°, allowing for accurate single crystal alignment with respect to magnetic field. In pulsed magnetic field experiments, accuracy of field alignment along Mg and B planes is estimated to be within 2-3°.

In Fig. 1, we present zero-field temperature dependence of the resistance for several $MgB_2$ single crystals. The inset in Fig. 1 is an enlarged view of the resistivity data near $T_c$. The results for three different samples demonstrate remarkable reproducibility giving a proof of high quality of our single crystals. In particular, all samples show sharp superconducting transitions at $T_c$, defined as the resistivity onset with criterion of 2% of the full resistivity drop, around 38.5-38.6K with a transition width $\Delta T(10\%-90\%)<0.3K$. Just above superconducting transition, at T=40K, we estimate $\rho=1\pm0.15\mu\Omega$ cm. Some scattering in the absolute values for the resistivity of different samples is believed to be due to the uncertainty in contact geometry resulting from small crystal size. In the normal state $\rho(T)$ dependence is well fitted to a power law $T^\alpha$ with $2.7<\alpha<2.8$ at T<200K, that is similar to previously reported results obtained on bulk sintered polycrystalline samples.[5,7] Also, for all three single crystals we found close values of residual resistivity ratio, RRR= $\rho$(273K)/$\rho$(40K)=4.9±0.3.

Characteristic results of our experiment are shown in Fig. 2. Plotted are resistive superconducting transitions for sample #1 at various magnetic fields up to 6T for H perpendicular to Mg and B planes (upper panel) and up to 9T in H//ab orientation (lower panel). In both orientations transport current direction is perpendicular to magnetic field. Before we start to describe anisotropic properties of $MgB_2$ single crystal, let us briefly consider peculiar voltage noise appearing in H//c field orientation in the low temperature part of superconducting transition at T<15K (see upper panel of Fig. 2). Although maximal value of noise level was found to decrease at smaller excitation currents (not shown), this noisy voltage response was clearly seen at the lowest value of transport current I=0.2mA (J≈4A/cm$^2$) used in our study. Very similar behavior was also found in the measurements on samples #2 and #3, thus, indicating its intrinsic origin, probably related to weak pinning properties of $MgB_2$ single crystal. However, for the context of the present paper, we note that this voltage noise does not affect our observation of anisotropic superconducting properties and defer its further discussion.

From different transition temperatures at a given value of magnetic field, anisotropic behavior of $MgB_2$ single crystal is clearly seen in Fig. 2. In particular, increasing H//c results in much more rapid $T_c$ decrease compared to H//ab field geometry. At H =6T superconducting transition is almost completely suppressed above T=4.2K, while in parallel field of the same magnitude it was found around 25K. Also, with the increase of magnetic field the structure of superconducting transition itself changes in a very different way in two field orientations. For H//ab the effect of magnetic field is to shift the transition to lower temperatures with a relatively small  T increase. On the contrary, in H//c orientation, at field of about 1-1.5T we found drastic change of the shape of superconducting transition. Fig. 3 illustrates this observation in more detail. Plotted are voltage response vs temperature for sample #3 taken at various currents at H=1T (upper panel) and H=2.5T (lower panel). At H=1T the superconducting transition is extremely sharp at low currents, while at H=2.5T the voltage response displays a more gradual behavior with temperature. This observation resembles transformation of superconducting transition near critical point in the melting line of "clean" $YBa_2Cu_3O_{7-}$ single crystals, where sharp resistance step associated with the flux line lattice melting below critical point was replaced by smooth continuous transition at higher fields.[17] From both panels in Fig. 3 one can also see strongly non-Ohmic behavior developing in the entire temperature region below $T_c$. We rule out a possibility that this non-linearity may be due to simple thermal effects since in measurements of zero-field transitions linear voltage response was found in the same range of excitation currents. We note, that non-Ohmic behavior found in $MgB_2$ single crystals is in striking contrast to linear voltage response observed in the broad region of vortex-liquid state above melting transition in $YBa_2Cu_3O_{7-}$ single crystals.[18,19]

Now we discuss magnetic phase diagram of $MgB_2$ single crystal deduced from our in-plane transport measurements (see Fig. 4). Since for sample #1 superconducting transition was measured by sweeping temperature at fixed magnetic fields, for each field three data points are shown as a function of temperature: resistivity onset, transition mid-point, and temperature of vanishing resistivity. On the other hand, for sample #2, measurements have been performed at fixed temperatures in a pulsed field, and the same three points on the superconducting transition are shown in dependence on field. As mentioned above, we associate resistivity onset with the upper critical field, $H_{c2}(T)$. The data for both single crystals demonstrate relatively close agreement. Slightly lower values of the upper critical fields for sample #2 may be due to possible field misalignment from the ab-planes in pulsed magnetic field experiment.[20] For both field orientations $H_{c2}(T)$ dependence shows distinct positive curvature extending from $T_c$ down to approximately 30K. Similar positive curvature of $H_{c2}(T)$ dependence was recently observed on dense $MgB_2$ wires[7] as well as sintered polycrystalline samples.[5,8] With the further decrease of temperature $H_{c2//}(T)$ as well as $H_{c2\perp}(T)$ dependence display approximately linear increase and below 15-20K start to saturate. Upper critical field anisotropy ratio, $\gamma = H_{c2//}/H_{c2\perp} = (m_c/m_{ab})^{0.5}$, calculated for sample #1 at different temperatures shows temperature dependence increasing from $\gamma = 2.2$ close to $T_c$ up to $\gamma \approx 3$ at T=30K. At lower temperatures $\gamma$ remains nearly unchanged. Obtained value of the electronic anisotropy $\gamma \approx 3$ places $MgB_2$ compound in between strongly anisotropic high-$T_c$ superconductors with anisotropy ratio ranging from $\gamma = 5$-7 for optimally doped $YBa_2Cu_3O_{7-\delta}$ up to $\gamma = 50$-200 for $Bi_2Sr_2CaCu_2O_{8+\delta}$ [21] and slightly anisotropic ($\gamma = 1$-1.15) other layered boride superconductor $RNi_2B_2C$ (R=Y, Lu) with crystal structure composed of alternating RC and $Ni_2B_2$ sheets.[22]

From the available data for sample #1 we estimate $H_{c2\perp}(0)$=7.0-7.5T and $H_{c2//}(0) = \gamma H_{c2\perp}(0)$=21-22T. According to the relations for anisotropic superconductors $H_{c2\perp}(T) = \Phi_0/(2\pi \xi_{ab}^2)$, where $\Phi_0$ is the flux quantum, and $\gamma = H_{c2//}(0)/H_{c2\perp}(0) = \xi_{ab}(0)/\xi_c(0)$ this corresponds to the in-plane and out-of-plane coherence length $\xi_{ab}(0) \approx 68$Å and $\xi_c(0) \approx 23$Å respectively. Using found in normal state resistivity measurements value $\rho(40K)=1\mu\Omega$ cm, the in-plane Fermi velocity of $4.9 \times 10^7$cm/s (Ref. 3), and a carrier density of $6.7 \times 10^{22}$ e/cm$^3$ for two free electrons per unit cell, we evaluate the in-plane electronic mean-free path of about 240Å near $T_c$. The out-of-plane resistivity for $MgB_2$ single crystals is about 4 times higher compared to the in-plane one.[23] Given the out-of-plane Fermi velocity of $4.76 \times 10^7$cm/s (Ref. 3) we get the out-of-plane mean-free path of about 60Å close to $T_c$. Thus, comparable values of coherence length and electronic mean-free path indicate our $MgB_2$ single crystal as approaching clean limit type-II superconductor.

Finally, we discuss peculiar upward curvature in $H_{c2}(T)$ dependence observed for $MgB_2$ single crystals near $T_c$. Similar positive curvature in $H_{c2}(T)$ has been previously reported for intermetallic borocarbides.[22,24,25] It was successfully described within effective two-band model for type-II superconductors in the clean limit.[24] As mentioned above, $MgB_2$ single crystal is also close to the clean limit. On the other hand, some difference between $MgB_2$ and borocarbides should be noted. In $RNi_2B_2C$ (R=Y, Lu) compounds $H_{c2}(T)$ dependence displays the same positive curvature for both parallel and perpendicular magnetic field orientations, and the out-of-plane anisotropy of $H_{c2}$ does not depend on temperature.[22,25] In striking contrast, in $MgB_2$ the upward curvature in $H_{c2}(T)$ is much more pronounced in parallel magnetic field orientation compared to perpendicular one (see inset in Fig. 4). In the temperature range above $0.7T/T_c$ $H_{c2}(T)$ dependence for both field orientations may be well fitted by the expression

$$H_{c2}(T)/H_{c2}(0)=A*(1-T/T_c)^\alpha, \quad (1)$$

with very different fitting parameters A and $\alpha$: A=1.225, $\alpha$=1.185 and A=1.62, $\alpha$=1.59 for H//c and H//ab correspondingly. Furthermore, due to the different temperature dependence of $H_{c2//}$ and $H_{c2\perp}$ the upper critical field anisotropy shows remarkable temperature dependence increasing from 2.2 close to $T_c$ up to about 3 below 30K. Also, it is worth mentioning that to the best of our knowledge exponent $\alpha$=1.59 obtained for H//ab geometry is one of the highest values so far experimentally found for any superconductor. From a theoretical point of view such large exponents exceeding 1.5 have been predicted for narrow-band systems with local attractive interactions.[26] However, a negative curvature of $H_{c2}(T)$ dependence clearly observed in $MgB_2$ at T<0.4-0.5$T_c$ (see Fig. 4) is different from the low-temperature behavior expected for superconductors with local pairing.[26] This brief discussion clearly demonstrates the need for more work on magnetic phase diagram of $MgB_2$ close to $T_c$ including detailed theoretical study of $H_{c2}(T)$ as well as further experiments on $MgB_2$ single crystals, e.g., measurements of $H_{c2}(T)$ by various methods.

In summary, we have used the in-plane transport measurements in magnetic field applied perpendicular and parallel to Mg and B sheets of $MgB_2$ single crystals to determine the upper critical field anisotropy of this recently discovered superconductor. The results indicate moderate value of the upper critical field anisotropy ratio that is nearly temperature independent at about 3.0 below 30K and monotonously decreases to =2.2 approaching $T_c$.

This work is supported by the New Energy and Industrial Technology Development Organization (NEDO) as Collaborative Research and Development of Fundamental Technologies for Superconductivity Applications.

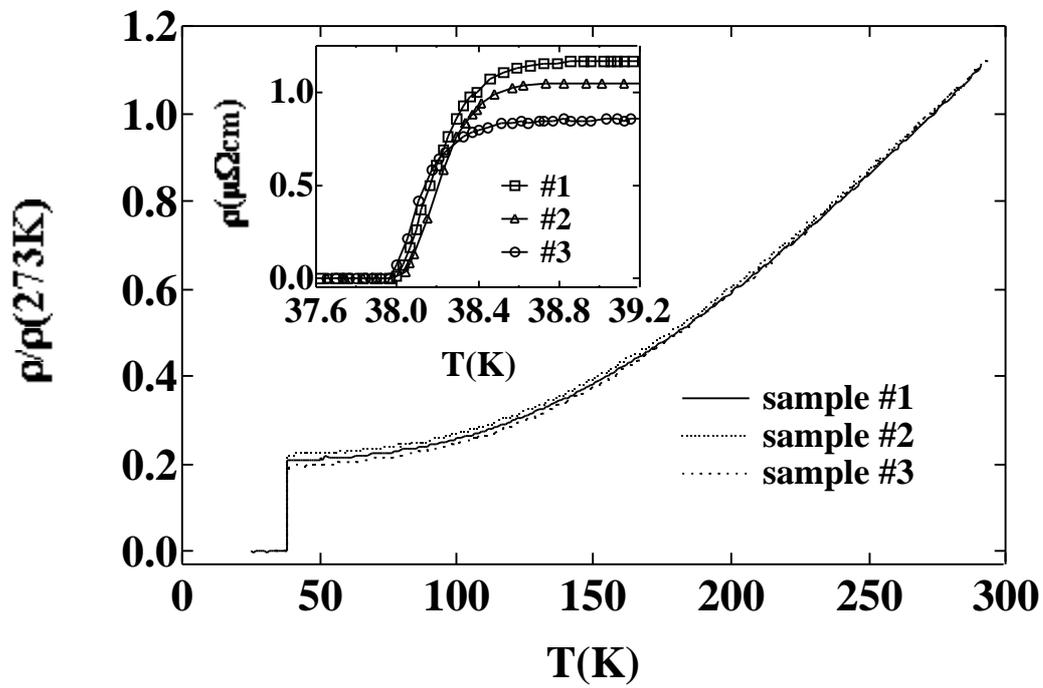

**Fig. 1.** Normalized zero-field temperature dependence of the in-plane resistivity for three different $MgB_2$ single crystals measured at current 0.5mA. Inset: Zero-field resistive superconducting transitions for the same crystals.

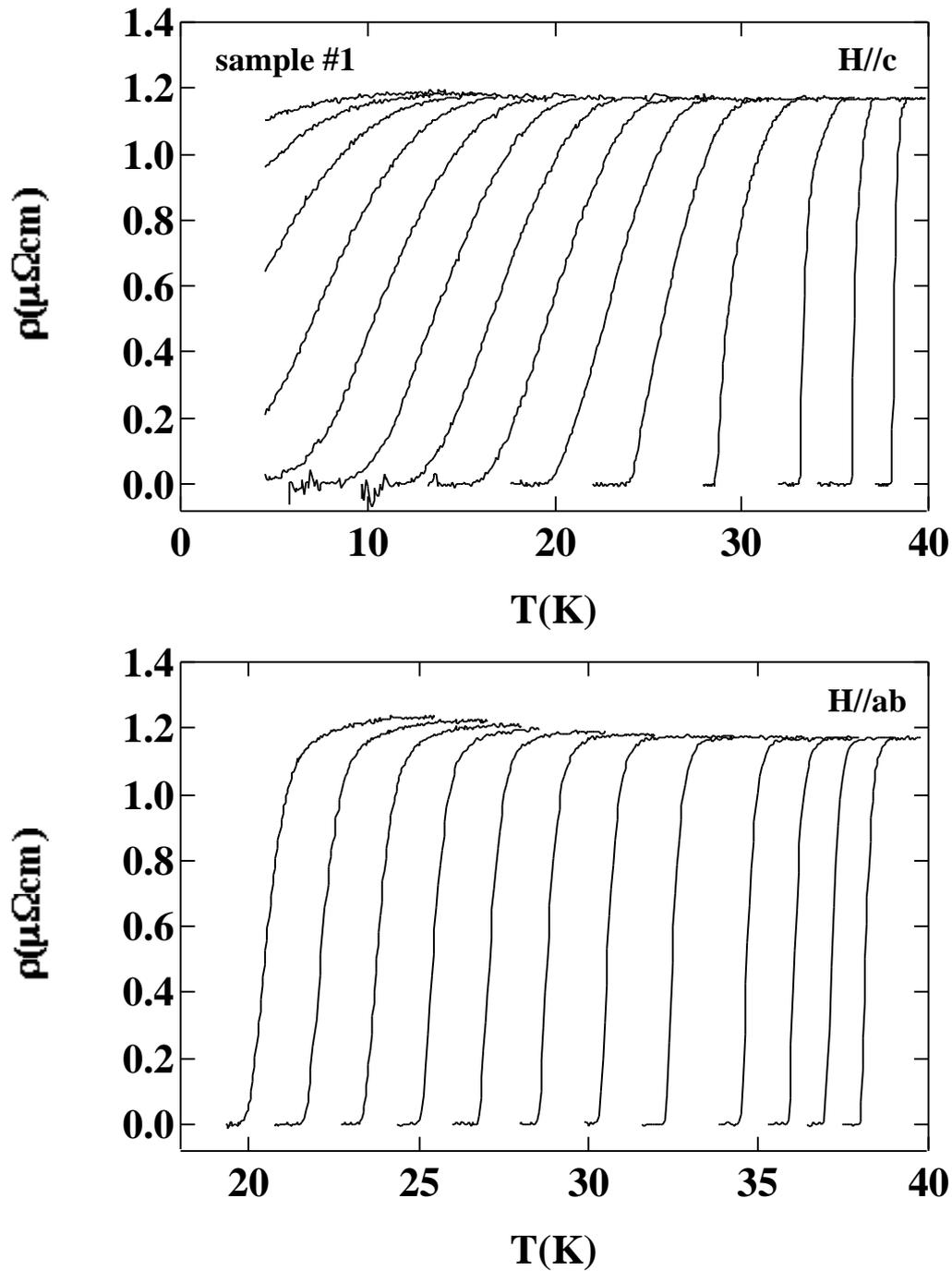

**Fig. 2.** Upper panel: Superconducting transitions at various magnetic fields of (from right to left) 0, 0.2, 0.5, 1, 1.5, 2, 2.5, 3, 3.5, 4, 4.5, 5, 5.5, and 6T applied perpendicular to Mg and B planes. Lower panel: Superconducting transitions in parallel fields of (from right to left) 0, 0.2, 0.5, 1, 2, 3, 4, 5, 6, 7, 8, and 9T. In both field orientations I=0.5mA and its direction is perpendicular to magnetic field. Different transition temperatures for a given applied field demonstrate the upper critical field anisotropy of $MgB_2$ single crystal.

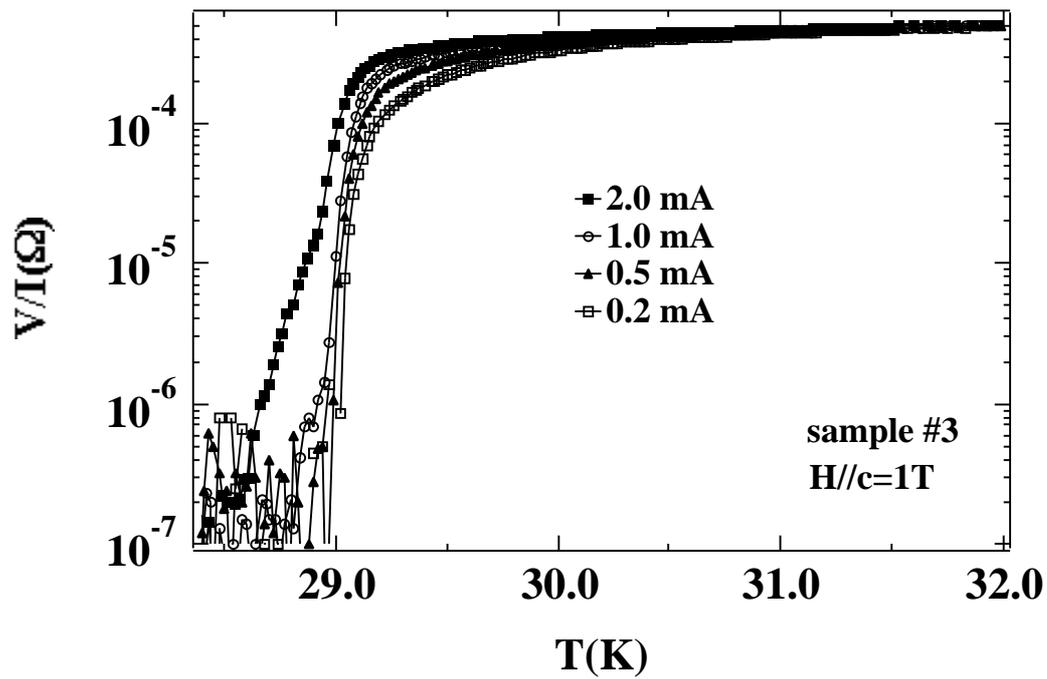

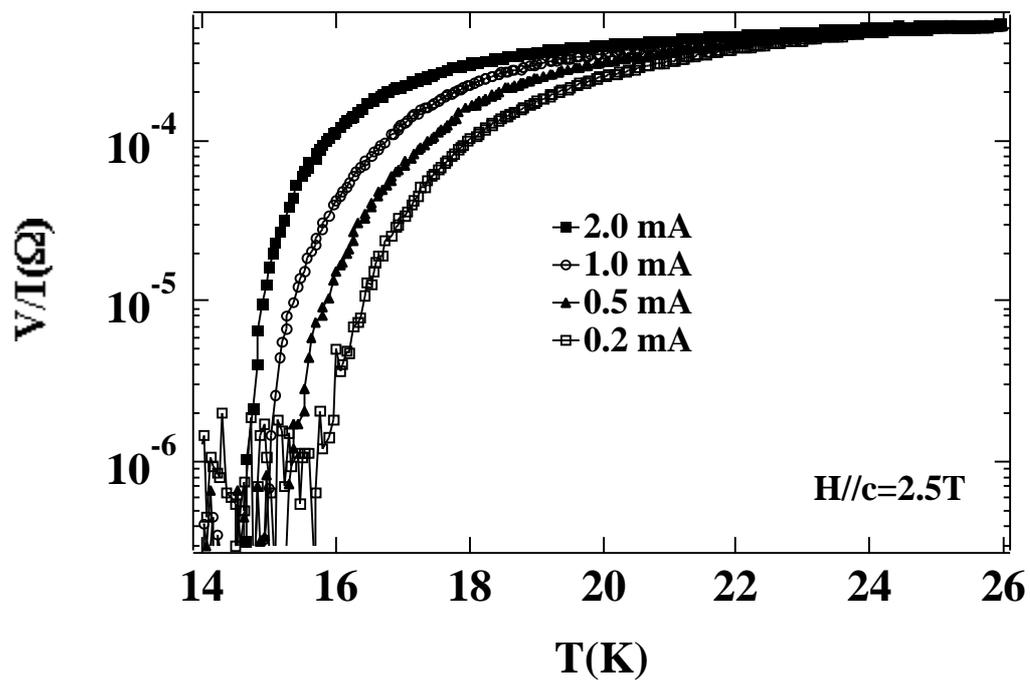

**Fig. 3.** Temperature dependence of the voltage response at various currents in magnetic field of 1T (upper panel) and 2.5T (lower panel) applied perpendicular to Mg and B planes.

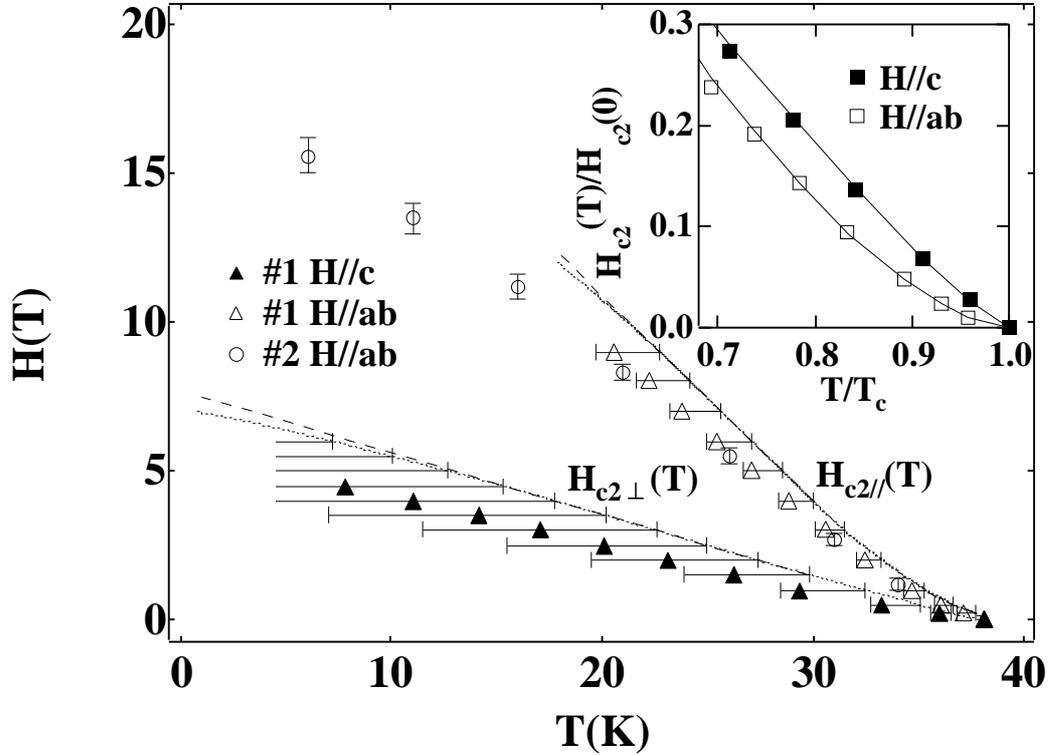

**Fig. 4.** Magnetic phase diagram of $MgB_2$ single crystal deduced from the in-plane temperature-dependent (sample #1) and field-dependent (sample #2) resistivity data as described in text. For sample #1 resistivity onset and transition end-point are vertical bars and symbols are mid-points of the transition as a function of temperature. For sample #2 horizontal bars and symbols show resistivity onset, completion and transition mid-point as a function of magnetic field correspondingly. For both samples the solid lines demonstrate transition width. The dotted lines are guides for eye and represent temperature dependence of parallel ($H_{c2//}$) and perpendicular ($H_{c2\perp}$) upper critical field for sample #1. The dashed lines represent a linear fit to $H_{c2//}(T)$ and $H_{c2\perp}(T)$ at intermediate temperatures. Inset: $H_{c2}(T)/H_{c2}(0)$ vs $T/T_c$ near $T_c$ for magnetic field parallel and perpendicular to Mg and B sheets ($T_c$=38.5K, $H_{c2//}(0)$=21T, $H_{c2\perp}(0)$=7.3T). Lines represent fit to Eqn. (1).